\documentclass[preprint]{vgtc}               %

\graphicspath{{figures/}{pictures/}{images/}{./}} %

\usepackage{times}                     %

\usepackage{tabu}                      %
\usepackage{booktabs}                  %
\usepackage{lipsum}                    %
\usepackage{mwe}                       %

\usepackage{mathptmx}                  %
\usepackage[dvipsnames]{xcolor}
\definecolor{blue}{HTML}{000000}

\onlineid{0}

\vgtccategory{Research}

\vgtcinsertpkg

\preprinttext{To appear in IEEE VIS 2025}

\title{Safire: Similarity Framework for Visualization Retrieval}

\author{Huyen N. Nguyen\orcid{0000-0001-6554-2327}\thanks{e-mail: huyen\_nguyen@hms.harvard.edu}\\ %
        \scriptsize Harvard Medical School %
\and Nils Gehlenborg\orcid{0000-0003-0327-8297}\thanks{e-mail: nils@hms.harvard.edu}\\ %
     \scriptsize Harvard Medical School}

\teaser{
  \vspace{-3mm}
  \centering
  \includegraphics[alt={A diagram illustrating a framework for visualization retrieval, split into two main sections: “Comparison Criteria (what?)” and “Representation Modalities (how?).” The top section, “Comparison Criteria,” contains two rows. The first row lists primary facets as blue boxes: Data, Visual Encoding, Interaction, Style, and Metadata. The second row, labeled “Derived properties,” has two green boxes: Data-centric Measure and Human-centric Measure. The bottom section, “Representation Modalities,” shows a two-dimensional grid with “Visualization Determinism” on the y-axis and “Information Content” on the x-axis. Three orange boxes—Raster Image, Vector Image, Specification—are higher on the grid, while a larger orange box labeled “Natural Language Description” sits lower on the y-axis but spans more of the x-axis. The figure caption explains that this framework distinguishes what aspects of visualizations are compared and how they are represented in retrieval systems, considering both information content and determinism of visualization rendering.}, width=0.6\linewidth]{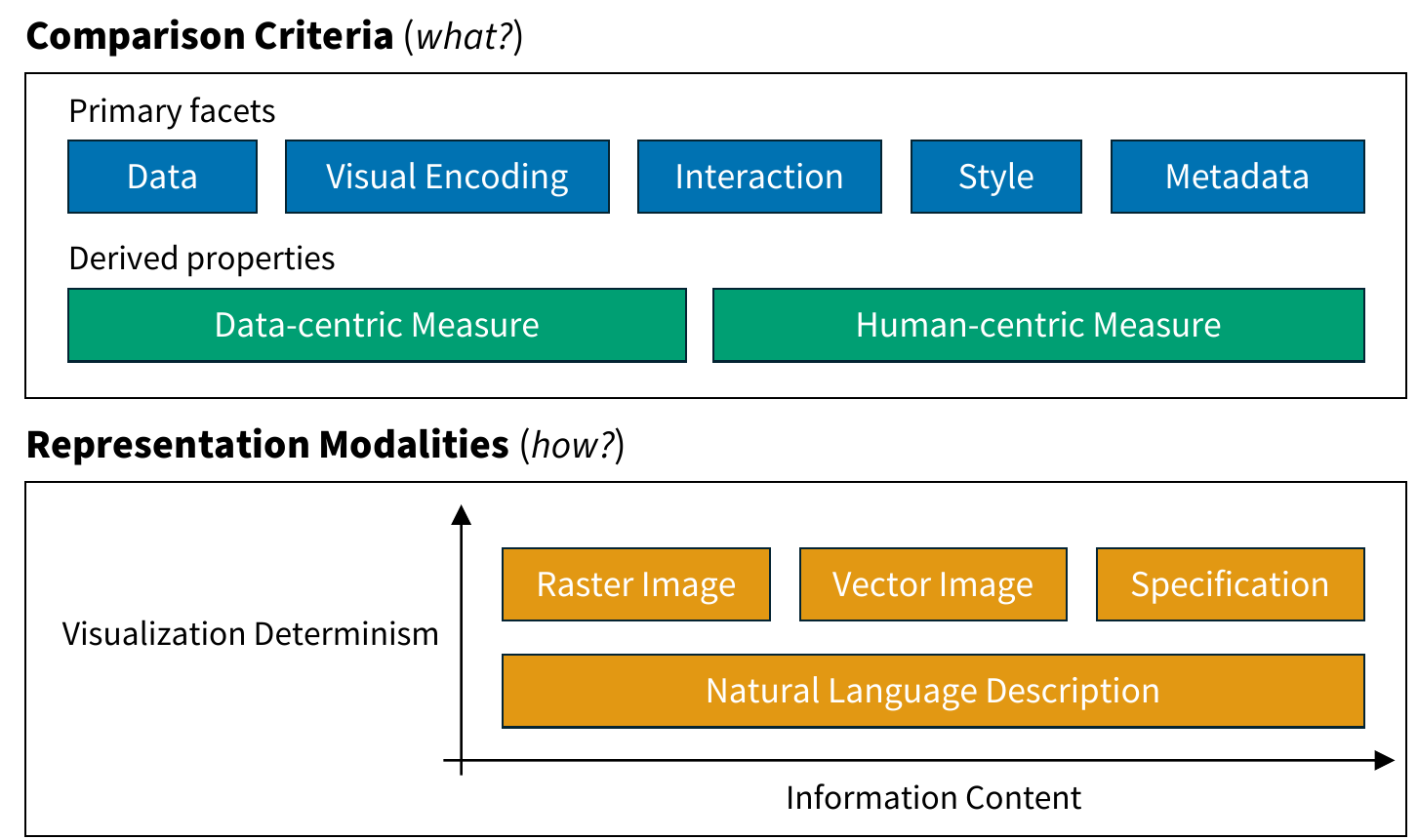}
  \vspace{-2mm}
  \caption{Our proposed similarity framework for visualization retrieval establishes clear comparison criteria and representation modalities.~The framework characterizes comparison criteria determining \textit{what} aspects of visualizations should be compared, while representation modalities define \textit{how} these visualizations are represented for comparison, with regard to information content and visualization determinism--the degree to which a representation format guarantees a single, consistent visual rendering. 
}
  \label{fig:teaser}
  \vspace{-0mm}
}

\abstract{Effective visualization retrieval necessitates a clear definition of similarity. Despite the growing body of work in specialized visualization retrieval systems, a systematic approach to understanding visualization similarity remains absent. We introduce the Similarity Framework for Visualization Retrieval (Safire), a conceptual model that frames visualization similarity along two dimensions: comparison criteria and representation modalities. Comparison criteria identify the aspects that make visualizations similar, which we divide into primary facets (data, visual encoding, interaction, style, metadata) and derived properties (data-centric and human-centric measures). Safire connects what to compare with how comparisons are executed through representation modalities. We categorize existing representation approaches into four groups based on their levels of information content and visualization determinism: raster image, vector image, specification, and natural language description, together guiding what is computable and comparable. We analyze several visualization retrieval systems using Safire to demonstrate its practical value in clarifying similarity considerations. Our findings reveal how particular criteria and modalities align across different use cases. Notably, the choice of representation modality is not only an implementation detail but also an important decision that shapes retrieval capabilities and limitations. Based on our analysis, we provide recommendations and discuss broader implications for multimodal learning, AI applications, and visualization reproducibility.
} %


\usepackage{svg}
\usepackage{float}
\newcommand{\orcid}[1]{\href{https://orcid.org/#1}{\includegraphics[width=10pt]{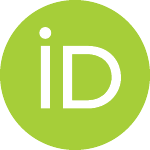}}}

\begin{document}

\firstsection{Introduction}

\maketitle

\label{sec:introduction}
Designing effective visualization retrieval systems involves unique challenges due to the distinctive nature of visualizations. While the overarching goal aligns with general information retrieval, which is to find relevant documents for a query, designers of visualization retrieval systems must first address what relevance means for visualizations. A fundamental question arises: \textit{What constitutes similarity between two visualizations?} This leads to a series of exploratory considerations: What criteria should guide comparison? Should we compare underlying data, visual encoding choices, interactive features, or aesthetic styles? Additionally, which representation format best captures a visualization's essence? These similarity modeling questions are critical in specialized visualization retrieval systems~\cite{hoque2019searching, nguyen2025geranium, oppermann2020vizcommender} and broader search platforms~\cite{chen2024chart2vec, xiao2023wytiwyr, ye2022visatlas}. Despite their recurrence across various scenarios, a systematic approach to clarifying essential dimensions of visualization similarity is currently lacking.

To address this gap, we propose a \underline{S}imil\underline{a}rity \underline{F}ramework for V\underline{i}sualization \underline{Re}trieval (Safire). Safire (pronounced similarly to sapphire) provides a structured framework for understanding visualization similarity along two key dimensions, as shown in Figure~\ref{fig:teaser}. The \textbf{comparison criteria} determine \textit{what} aspects of visualizations should be compared, while \textbf{representation modalities} define \textit{how} visualizations are represented for comparison. We ground Safire in visualization theory and contextualize it with practical applications to ensure its applicability in real-world systems.

We develop criteria for what makes visualizations similar, distinguishing between primary facets used in visualization construction and derived properties observed afterward. The framework identifies five key primary facets: data, visual encoding, interaction, style, and metadata, drawing from both visualization theory and practical system needs. Derived properties cover both data-centric computational metrics and human-centric perceptual aspects.

The framework connects what to compare with how comparisons are performed through representation modalities. 
Appropriate representation forms the basis of effective retrieval, and this principle applies to visualization retrieval as well. The chosen representation format (e.g., declarative specification, raster image) dictates which aspects are captured and which similarity criteria are accessible for comparison. Based on the information content and visualization determinism, we categorize the existing representation modalities into four groups: raster image, vector image, specification, and natural language description.

We analyze several visualization retrieval systems using Safire to demonstrate its practical value in clarifying criteria and modalities. We find that the choice of representation is not only an implementation detail but a decision that shapes the possibilities and limitations of the retrieval process. We then provide recommendations and discuss implications in the bigger context of retrieval, multimodal learning, AI applications, and reproducibility. Our contributions are two-fold:

\vspace{-1mm}

\begin{itemize}
    \item A similarity framework for visualization retrieval, Safire, emphasizing comparison criteria and representation modalities. This conceptual model serves as a practical guide for system builders to clarify their design choices and select similarity dimensions that align with their intended use cases.
    \vspace{-1mm}
    \item An application of Safire to analyze existing visualization retrieval systems, highlighting different solution patterns in current approaches, from which we discuss broader implications for retrieval, reproducibility, and AI applications.

\end{itemize}

\vspace{-3mm}

\section{Related Work}
\label{sec:related_work}

The formulation of Safire as a framework was inspired by how the nested model~\cite{munzner2009nested} frames different facets of visualization design, along with its extension~\cite{meyer2012four} for inter- and intra-level blocks. FaEvR~\cite{vaidya2020knowing} provides an exemplar model that gathers insights from real-world visualizations to build a framework, and then applies this framework to analyze these visualizations from a different angle. 
Although visualization retrieval has unique characteristics inherent to the visual representation of data, it shares a common goal with image and other types of information retrieval~\cite{bannour2013building}: 
to find relevant documents that match a query conveying an information need~\cite{IRbook}.

Building on these foundational ideas, our work is informed by insights from prior visualization retrieval systems, which have highlighted approaches for modeling similarity in visualizations~\cite{chen2024chart2vec, hoque2019searching, li2022structure, oppermann2020vizcommender, setlur2023olio, xiao2023wytiwyr, ye2022visatlas, ying2024vaid}. Existing systems typically address only a subset of possible criteria, with each focusing on different aspects. For example, ChartSeer~\cite{zhao2020chartseer} primarily considers visual representation and data variables for chart summarization. In contrast, Safire introduces a unifying abstraction of \textit{primary facets} that spans five dimensions: data, visual encoding, interaction, style, and metadata, providing a more comprehensive framework than any single previous work. Our framework further complements this with a parallel concept of \textit{derived properties}, together forming a comprehensive model for framing similarity.
For the representation modalities in Safire, we reference the visualization workflow using D3~\cite{D3}: from imperative programming, to vector graphics (SVG) with interactions~\cite{wordstream, Malview, maker}, to vector/raster image export~\cite{RAWGraphs}.  These modalities will be described in greater detail in the following sections.

\section{Safire: Similarity Framework for Visualization Retrieval}
\label{sec:framework}

\begin{figure*}[t]
 \centering 
 \includegraphics[alt={A diagram illustrating four modalities for representing a bar chart visualization: raster image (PNG), vector image (SVG), specification (JSON), and natural language description. On the left, a PNG bar chart shows annual values from 2020 to 2024, with the bar for 2021 as the maximum. To its right, an SVG markup snippet is shown, and next to it is the rendered SVG chart (identical to the PNG), labeled as "SVG." Further right is a JSON specification for the chart using Vega-Lite schema, labeled "JSON." Below these, orange boxes labeled "Raster Image," "Vector Image," and "Specification" are displayed horizontally and connected by arrows indicating 1:1 mapping relationships. Underneath, a larger orange box labeled "Natural Language Description" is linked to each of the other three boxes via arrows labeled "1..n," representing multiple possible interpretations. Examples of detailed caption and natural language descriptions are shown below the diagram. The figure demonstrates how visualizations can be represented across specification, vector, raster, and textual modalities, with 1:1 mapping between image and specification formats, and more flexible mappings for natural language.},width=.8\linewidth]{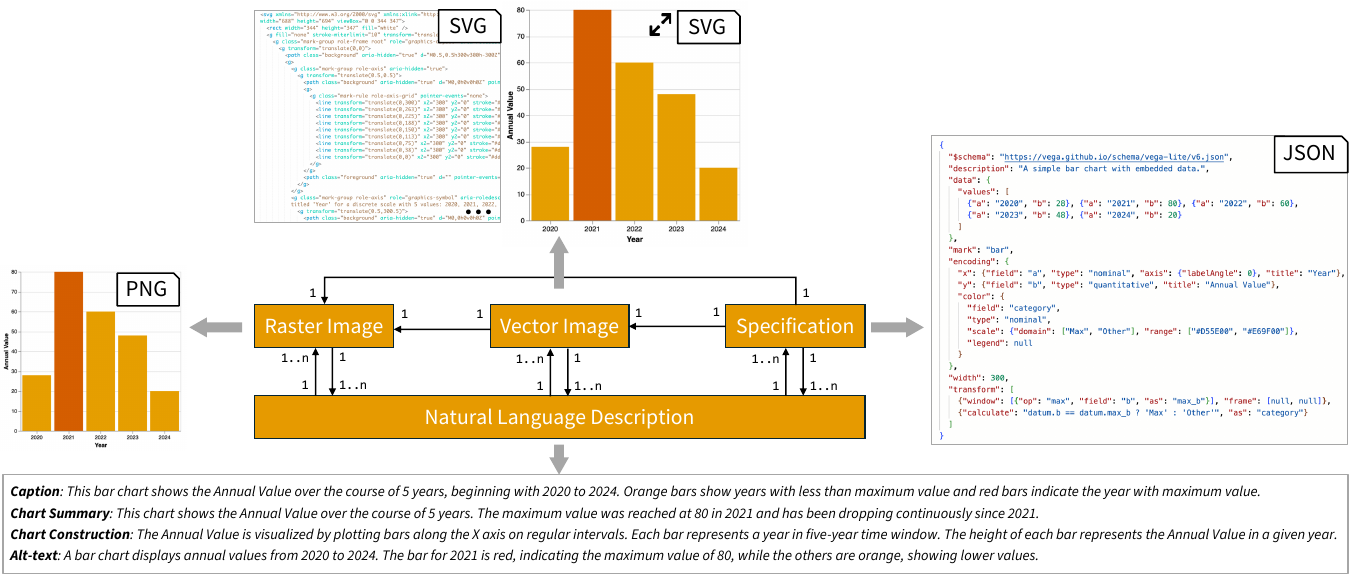}
 \vspace{-2mm}
 \caption{Visualization representation across four modalities: a Vega-Lite JSON specification (right) rendered as SVG vector--with accompanying SVG markup, and PNG raster images, along with multiple natural language descriptions. Specification, vector, and raster formats maintain 1:1 mapping relationships (directed arrows), while natural language enables one-to-many interpretations (multiple text examples).}
 \label{fig:example}
\vspace{-5mm}
\end{figure*}

\subsection{Comparison Criteria}

In our framework, the criteria answer the question of `what’ aspects should be compared for understanding similarity between different visualizations. As presented in the top panel of Figure~\ref{fig:teaser}, we distinguish primary facets that directly contribute to constructing a visualization from derived properties that are extracted after a visualization is built. This distinction acknowledges the fundamental difference between the contributing parameters that define how a visualization is created and the emergent characteristics that can only be observed in the final visual output. The following sections elaborate on how we developed these criteria.

\subsubsection{Primary Facets}

Our framework integrates criteria from theoretical visualization models and empirical retrieval systems, resulting in a unified five-facet approach that provides more comprehensive coverage than previous systems~\cite{chen2024chart2vec, hoque2019searching, li2022structure, oppermann2020vizcommender, setlur2023olio, xiao2023wytiwyr, ye2022visatlas, ying2024vaid,zhao2020chartseer}.
We ground our approach in fundamental models of visualization design, particularly the nested model by Munzner~\cite{munzner2009nested} and its subsequent extensions to inter- and intra-level blocks by Meyer et al.~\cite{meyer2012four}. These models systematically deconstruct visualization design into core elements, conceptualizing visualization creation as a cascade of decisions that transform domain problems into data-task abstractions, visual encodings, and implementations.

By analyzing these distinct design layers, we identify the first two fundamental comparison groups: (1) underlying \textbf{data} and (2) \textbf{visual encoding} that maps data attributes to visual features. 
Given the increasing importance of interactivity in visualization workflows~\cite{dimara2019interaction, franconeri2021science, nguyen2021interactive}, we deem it only appropriate to include (3) \textbf{interaction} as a separate dimension focused on user-centric exploration.
Observations from practical visualization retrieval systems and broader design considerations~\cite{hoque2019searching, xiao2023wytiwyr, setlur2023olio, nguyen2025geranium, saleh2015learning} emphasize the importance of including (4) \textbf{visual styles} and (5) \textbf{contextual metadata} as additional criteria. The criteria are defined as follows:

\paragraph{\textbf{Data}} Covers data-related properties, including transformation methods, parsing, data types, and aggregation parameters (e.g., binning size). This criterion facilitates searching for visualization examples handling specific data types or wrangling approaches.

\vspace{-.4mm}

\paragraph{\textbf{Visual Encoding}} Represents the mapping of data to visual attributes, such as mark types, layout structures, and visual channels to encode values (e.g., bar height, circle radius). This criterion enables identification of visually similar representations, such as bar charts using bar length to indicate the magnitude of a value.

\vspace{-.4mm}

\paragraph{\textbf{Interaction}} Captures user interactivity with visual elements, including brushing, linking, and details-on-demand features. This criterion supports exploration of interactive techniques, e.g., linking an overview with detailed views following user selection.

\vspace{-.4mm}

\paragraph{\textbf{Style}} Corresponds to non-data-encoding visual attributes~\cite{hoque2019searching} that contribute to aesthetic and perceptual aspects, including typography, background colors, and decorative elements. This criterion facilitates discovery of visual language applications, e.g., similar color palette usage across different contexts.

\vspace{-.4mm}

\paragraph{\textbf{Metadata}} Comprises information that describes and contextualizes the visualization, including titles, subtitles, legends, and annotations. This criterion supports identification of effective approaches for enhancing visualization comprehensibility through supplementary elements.

It is important to note that these five primary facets are not mutually exclusive. Depending on the specific domain problem and task, an attribute can belong to multiple categories. For example, stroke width can be a visual encoding when it corresponds to value magnitude, or style when its purpose is to enhance legibility.

\subsubsection{Derived Properties}

Having established primary facets that define visualization construction, our framework now addresses derived properties: features extracted or computed from the resulting visualization. This characterization aligns with the role of visualization in visual analytics (VA) workflows: providing the means for communicating about data and information, where humans and machines cooperate~\cite{keim2008visual}. Inspired by the systematic considerations in VA by Sun et al.~\cite{sun2022toward}, we divide derived properties into two categories:

\paragraph{\textbf{Data-centric Measure}} Refers to computational properties derived from data, designed for analytical interpretation. Examples include distribution, outliers, and cluster-related measures~\cite{ying2024vaid}. This criterion enables finding visualizations with specific computational targets, topologies, or statistical measures.

\paragraph{\textbf{Human-centric Measure}} Characterizes how users perceive information, involving human cognitive processing of visual information. Examples include metrics for perceptual similarity~\cite{pandey2016towards}, reflecting how observers group plots based on concepts like orientation, edges, or density. This criterion supports identifying visualizations grouped based on human perceptual judgments.

\subsection{Representation Modalities}

Representation modality defines how visualization information is represented. Before creating vector embeddings as the computable and comparable format, it is essential to characterize raw modalities that capture different aspects of the visualization (Figure~\ref{fig:example}). Common modalities include raster images (PNG, JPG file formats), vector graphics (SVG), and declarative specifications (JSON).

We categorize the raw modalities along two dimensions: information content and visualization determinism (Figure~\ref{fig:teaser}). Higher information content enables users to recreate the visualization more accurately and extract more meaningful information. Visualization determinism refers to the degree to which a representation format guarantees a singular, consistent visual rendering without requiring additional interpretation. These two dimensions are essential for retrieval due to their immediate association with how much information is captured and how consistently that information translates to a specific visual form. We define the representations as follows:

\paragraph{\textbf{Raster Image}}  Renders a visualization as a fixed grid of pixels (e.g., PNG, JPG). Each pixel stores only color information without preserving data relationships or visual mark semantics. As a raw modality for visualization retrieval, raster images require visual feature extraction via a predefined taxonomy or deep learning models to interpret chart types~\cite{ye2022visatlas, xiao2023wytiwyr}. While suitable for image-based retrieval or search-by-sketch scenarios, they lack structural relationships to the underlying data.

\paragraph{\textbf{Vector Image}} Preserves visualization geometry through scalable paths, shapes, and text elements that can be scaled without loss of quality. Examples include an SVG file of a scatter plot that represents each point as a circle with properties like position, radius, and color. SVG uses HTML-tag markup that, along with its visual rendering, can enable structure-aware retrieval~\cite{li2022structure}.

\paragraph{\textbf{Specification}} Defines the visualization's structure, data bindings, encoding rules, and potentially interaction, at a high level with a predefined schema. Specifications offer machine-readable access to high-level semantics. They are ideal for precise matching and retrieval based on structural similarity or query-by-example, including searching for \textit{interaction}. Examples include retrieval systems Chart2Vec~\cite{chen2024chart2vec} and Geranium~\cite{nguyen2025geranium} (JSON format), and recommender system VizCommender~\cite{oppermann2020vizcommender} (Tableau Workbook XML).

\paragraph{\textbf{Natural Language (NL) Description}} Captures the semantic content of a visualization using NL to convey and contextualize insights. Examples include alt-text, which is the most abstract, human-readable interpretation of the visualization~\cite{smits2024altgosling, smits2024explaining}. Other examples are captions (general interpretation), chart summaries (richer descriptions of patterns, insights, and context, but may lack encoding information), and chart construction (procedural instructions for building charts--similar to grammar-based specification but in NL). NL descriptions inherently contain ambiguity: visualizations can have multiple descriptions for different audiences, and different charts of the same data may deliver a similar message.

Figure~\ref{fig:example} demonstrates the interconnections between these modalities. A Vega-Lite JSON specification defines the visualization structure, rendered as an SVG vector image (along with its markup) and captured as a PNG raster image. While specification, vector, and raster representations maintain a 1:1 mapping (along directed arrows), NL descriptions exhibit one-to-many relationships, as shown by the four different textual representation types: caption, chart summary, chart construction, and alt-text.

\section{Application Examples}
\label{sec:examples}

In this section, we analyze several existing visualization retrieval systems using our Safire framework. These applications provide contexts for how the visualization retrieval problem can be approached in different usage scenarios. We enhance our examples by outlining each solution pattern in terms of Safire's vocabulary, allowing system builders to systematically review the choices of criteria and modalities.

\subsection{Searching D3 Visualizations}

\begin{figure}[H]
 \centering 
 \includegraphics[width=.7\linewidth]{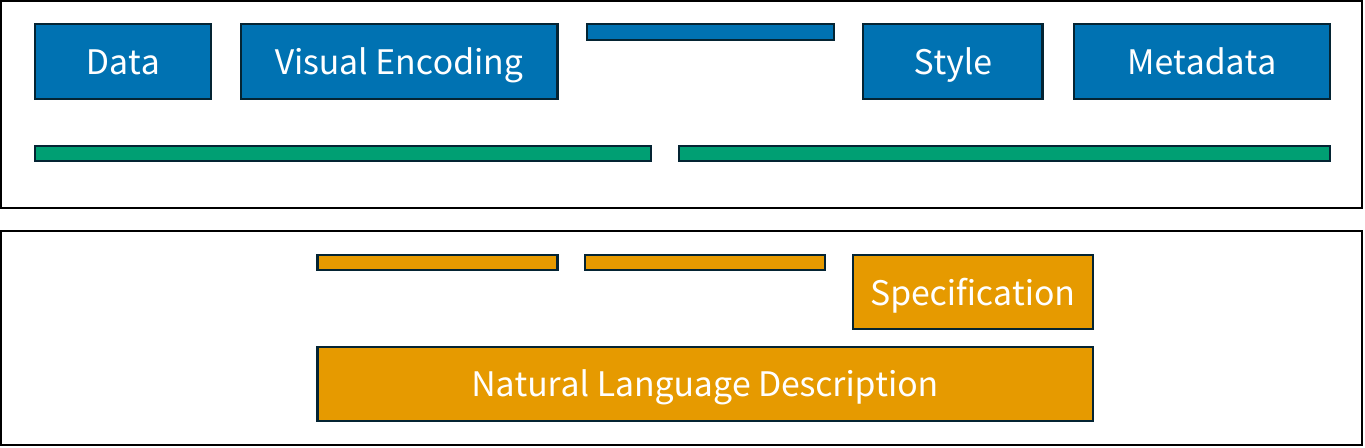}
\vspace{-1mm}
 \caption{Searching D3 Visualizations~\cite{hoque2019searching}}
 \label{fig:searchingD3}
\vspace{-2mm}
\end{figure}

Hoque and Agrawala present a system for searching D3 visualizations by visual style and structure~\cite{hoque2019searching}, as shown in Figure~\ref{fig:searchingD3}. Their retrieval system deconstructs and indexes visualizations based on data, visual encoding, style, and metadata criteria. The system generates a representation similar to a Vega-Lite~\cite{satyanarayan2016vega} specification for each visualization, which also serves as the query input format. NL text and metadata are indexed separately alongside the deconstructed specification. This work demonstrates the flexibility of specification in encoding chart semantics. By extracting both data- and non-data-encoding attributes, this approach enables comprehensive searches across visual and structural dimensions, even with partial specifications.

\subsection{Multimodal Retrieval of Genomics Visualizations}
\label{sub:sec:mm}
Nguyen et al.~\cite{nguyen2025geranium} present a multimodal retrieval system for genomics data visualizations, covering all five comparison criteria: data, visual encoding, interaction, style, and metadata. Their system uses three modalities: raster images, Gosling~\cite{lyi2021gosling} grammar specifications, and NL descriptions (both alt-text and LLM-enriched versions).

\vspace{-2mm}
\begin{figure}[H]
 \centering 
 \includegraphics[width=.7\linewidth]{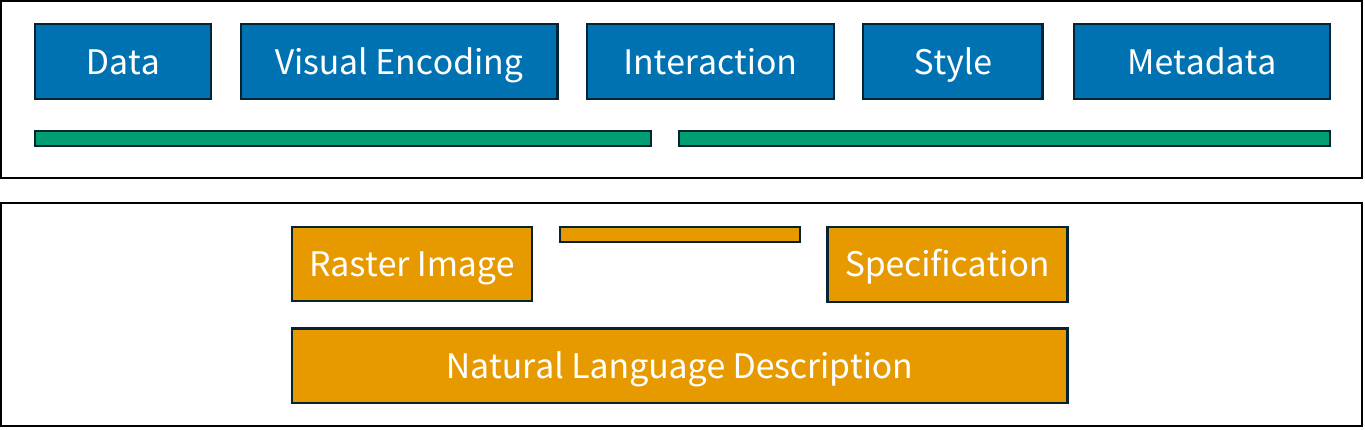}
 \vspace{-1mm}
 \caption{Multimodal Retrieval of Genomics Data Visualizations~\cite{nguyen2025geranium}}
 \label{fig:geranium}
\vspace{-2mm}
\end{figure}

The multimodal representations approach enables the system to capture both the semantic structure and visual characteristics of genomics visualizations, supporting flexible querying by example images, text queries, or specification-based queries. 

\subsection{WYTIWYR: User Intent-Aware Framework}

Xiao et al. present WYTIWYR~\cite{xiao2023wytiwyr}, a retrieval tool that compares charts based on visual attributes and style cues. To better understand user intent, the authors first conducted a preliminary study to formulate chart attributes along three dimensions: colormap, data trends, and view layout.

\vspace{-2mm}
\begin{figure}[H]
 \centering 
 \includegraphics[width=.7\linewidth]{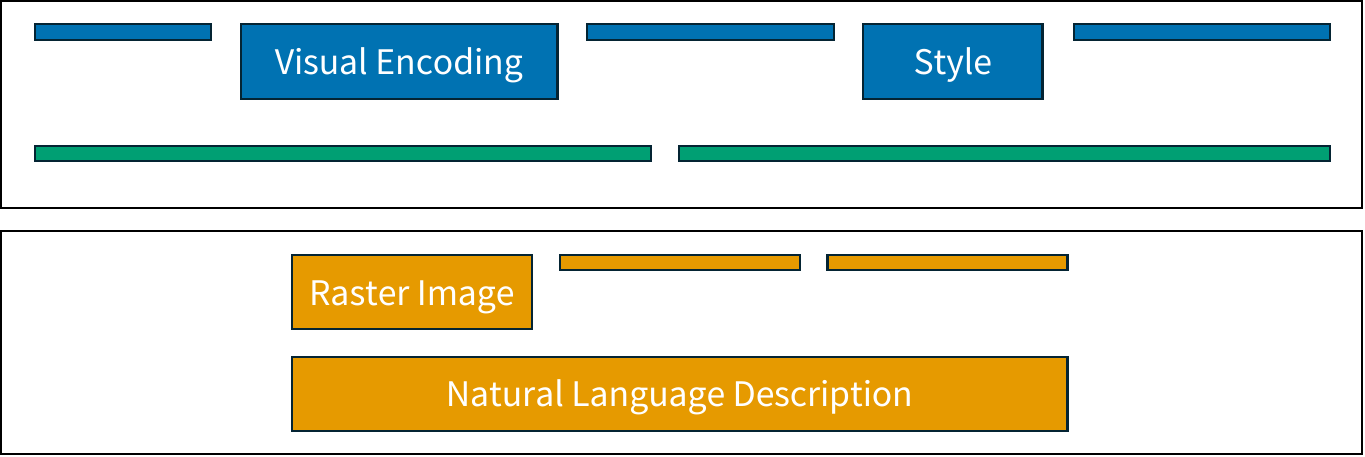}
 \caption{WYTIWYR: User Intent-Aware Framework~\cite{xiao2023wytiwyr}}
 \label{fig:wytiwyr}
\vspace{-2mm}
\end{figure}

The system processes raster images as visualization inputs, with optional text prompts expressing user intent, and combines them via a CLIP-based multimodal encoder. 

\subsection{VAID: Indexing View Designs in VA system}
\label{sub:sec:vaid}
Ying et al. present VAID~\cite{ying2024vaid}, an index structure for complex and composite visualizations. VAID compares both primary facets (data-related, visual encoding, and style) and derived data-centric measures: graph-related metrics (e.g., clusters, topology) and tabular structures (e.g., correlation, distribution, outliers).

\vspace{-2mm}
\begin{figure}[H]
 \centering 
 \includegraphics[width=.7\linewidth]{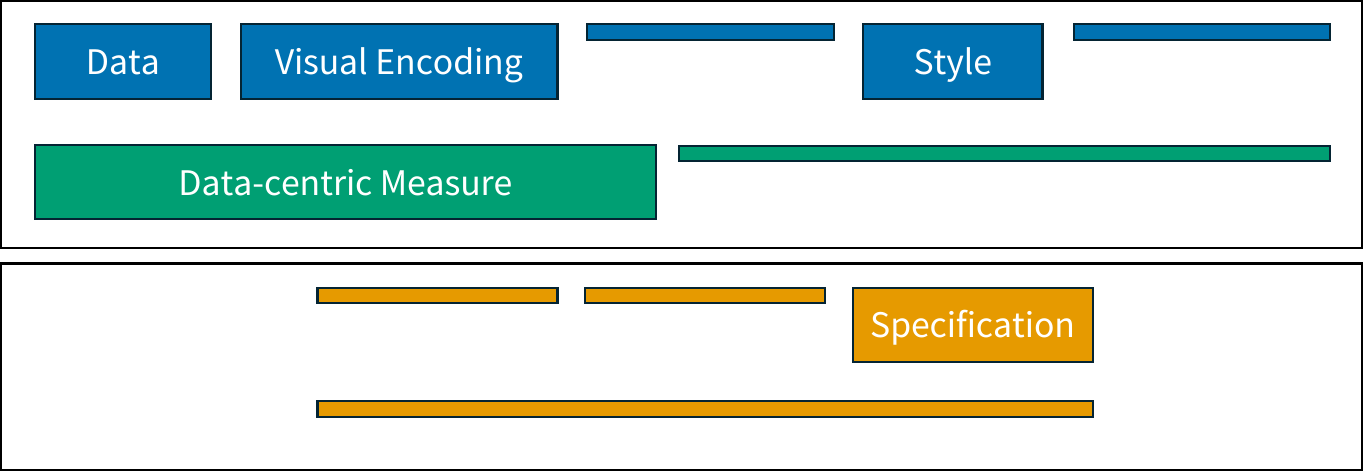}
 \caption{VAID: Indexing View Designs in VA system~\cite{ying2024vaid}}
 \label{fig:vaid}
\end{figure}
\vspace{-2mm}
Although VAID provides multiple criteria for comparison, it indexes views solely through specifications, using an extended Vega-Lite grammar, demonstrating the comprehensiveness of specification-based representation.

\section{Discussion}
\label{sec:discussion}

\paragraph{Representation Modality Shapes Retrieval Capabilities and Reproducibility.} We find that data- and interaction-related criteria are only comparable when specification is involved. In fact, specification is one of the most versatile modalities, encompassing all five primary facets  (Section~\ref{sub:sec:mm}) and multiple data-centric measures (Section~\ref{sub:sec:vaid}). 
Vector images integrate benefits from both raster images and specifications but often feature complex, highly nested markup. Meanwhile, NL descriptions can capture high-level insights and context missing in other modalities, yet their inherent ambiguity challenges precise matching and retrieval. Recognizing these trade-offs, multimodal retrieval presents a promising approach that integrates complementary strengths of each modality to create a more comprehensive understanding. 
From our observations and application examples, we note that both data-centric and human-centric measures are still under development in this space, with limited work applying these criteria in retrieval.
In terms of reproducibility and information content, specifications rank highest, followed by vector images and then raster images. This aspect is essential for visualization authoring~\cite{van2024understanding}, where retrieved examples can serve as both inspiration and templates for adaptation. Specifications enable efficient programmatic modifications, while raster images serve as strong visual references but with limited editability.

\paragraph{NL Description Is Highly Nondeterministic.} In contrast to 1:1 mappings in specifications and vector images, NL descriptions are inherently ambiguous, resulting in one-to-many relationships with visualizations. Within the Safire framework, NL description therefore exhibits low visualization determinism, and its information content varies greatly by description type.
The same chart can be described in various ways: some descriptions specify data bindings and encodings, while others focus on broader patterns or insights. Specifically, a \textit{chart construction} involves procedural instructions, functioning as specifications written in NL rather than in a formal grammar. In contrast, a \textit{chart summary} can convey insights beyond the visual channel, such as mark type.
Here, visual features serve merely as the medium to extract meaning. These observations complement the four-level model of semantic content~\cite{lundgard2021accessible} by considering the nuanced nature of descriptions, which varies with communication intent and context. 
NL descriptions associated with visualizations thus present a rich direction for further investigation.

\paragraph{Guidance for LLMs in AI Applications.} The five primary facets of visualization can help guide large language models (LLMs) to focus on key elements and steer their interpretation of charts toward clearer, more accurate understanding. By structuring prompts around data, visual encoding, interaction, style, and metadata, we can direct the LLMs' attention to areas they might otherwise overlook. Furthermore, these facets create a systematic way to evaluate LLM performance in visualization comprehension tasks, revealing which aspects remain challenging and may require additional prompt engineering or model training for improvement.

\section{Conclusion and Future Work}
\label{sec:conclusion}

We introduced Safire, a framework for modeling visualization similarity that connects comparison criteria with representation modalities. Safire offers a structured approach to defining similarity across modalities, each with different implications for retrieval, comprehension, and reproducibility. Applying Safire to existing retrieval systems demonstrated its value in outlining design decisions and aligning similarity dimensions with intended use cases. While algorithm comparison is beyond our scope, prior work on formal evaluation suggests promising directions. 
In future work, we will evaluate the feasibility of Safire with users who build and design retrieval systems.
Additionally, we plan to evaluate Safire with leading LLM-based retrieval methods to highlight its effectiveness and to examine the strengths and limitations of different approaches.

\acknowledgments{
This work was supported in part by the National Institutes of Health (R01HG011773) and the Advanced Research Projects Agency for Health (AY2AX000028).}

\bibliographystyle{abbrv-doi}
\bibliography{references}
\end{document}